\pdfoutput = 1

\documentclass[%
 reprint,
superscriptaddress,
twocolumn,
 amsmath, amssymb,
 aps,
 prl,
]{revtex4-1}

\usepackage[utf8]{inputenc}
\usepackage[T1]{fontenc}

\usepackage{mathrsfs}
\usepackage{graphicx}
\usepackage{bm}
\usepackage{float}
\usepackage[colorlinks = true,
            linkcolor = blue,
            urlcolor  = blue,
            citecolor = blue,
            anchorcolor = black]{hyperref}

\usepackage{pgffor}

\newcommand{\TA}[1]{{\color[rgb]{0,0.,0.}#1}}

\begin{document}

\preprint{APS/123-QED}

\title{State equation from the spectral structure of human brain activity}






\author{Trang-Anh Nghiem}
\thanks{These authors contributed equally.}
\affiliation{European Institute for Theoretical Neuroscience, 74 Rue du Faubourg Saint-Antoine, Paris, 75012 France}
\affiliation{Unit of Neuroscience, Information, and Complexity, Centre National de la Recherche Scientifique, 1 Avenue de la Terrasse, Gif-sur-Yvette 91190 France}
\author{Jean-Marc Lina}
\thanks{These authors contributed equally.}
\affiliation{Centre de recherches math\'ematiques, Universit\'e de Montr\'eal, 2920, Chemin de la tour, bur. 5357 Montréal, Québec, H3T 1J4, Canada}
\affiliation{Département de génie électrique, École de technologie supérieure, 1100 rue Notre-Dame Ouest, Montréal, Québec, H3C 1K3 Canada}
\author{Matteo di Volo}
\affiliation{European Institute for Theoretical Neuroscience, 74 Rue du Faubourg Saint-Antoine, Paris, 75012 France}
\affiliation{Unit of Neuroscience, Information, and Complexity, Centre National de la Recherche Scientifique, 1 Avenue de la Terrasse, Gif-sur-Yvette 91190 France}
\author{Cristiano Capone}
\affiliation{European Institute for Theoretical Neuroscience, 74 Rue du Faubourg Saint-Antoine, Paris, 75012 France}
\affiliation{Unit of Neuroscience, Information, and Complexity, Centre National de la Recherche Scientifique, 1 Avenue de la Terrasse, Gif-sur-Yvette 91190 France}
\author{Alan C. Evans}
\affiliation{Ludmer Centre for Neuroinformatics and Mental Health, Montreal Neurological Institute and Hospital, 3801 Rue University, Montréal, Quebec, H3A 2B4 Canada}
\affiliation{Department of Biomedical Engineering, McGill University, 3775 Rue University, 316, Montréal, Quebec, H3A 2B4 Canada }
\author{Alain Destexhe}
\thanks{These authors contributed equally. }
\affiliation{European Institute for Theoretical Neuroscience, 74 Rue du Faubourg Saint-Antoine, Paris, 75012 France}
\affiliation{Unit of Neuroscience, Information, and Complexity, Centre National de la Recherche Scientifique, 1 Avenue de la Terrasse, Gif-sur-Yvette 91190 France}
\author{Jennifer S. Goldman}
\thanks{These authors contributed equally. }
\affiliation{European Institute for Theoretical Neuroscience, 74 Rue du Faubourg Saint-Antoine, Paris, 75012 France}
\affiliation{Unit of Neuroscience, Information, and Complexity, Centre National de la Recherche Scientifique, 1 Avenue de la Terrasse, Gif-sur-Yvette 91190 France}
\affiliation{Centre de recherches math\'ematiques, Universit\'e de Montr\'eal, 2920, Chemin de la tour, bur. 5357 Montréal, Québec, H3T 1J4, Canada}
\affiliation{Ludmer Centre for Neuroinformatics and Mental Health, Montreal Neurological Institute and Hospital, 3801 Rue University, Montréal, Quebec, H3A 2B4 Canada}

\maketitle
\textbf{Neural electromagnetic (EM) signals recorded non-invasively from individual human subjects vary in complexity and magnitude. Nonetheless, variation in neural activity has been difficult to quantify and interpret, due to complex, broad-band features in the frequency domain. Studying signals recorded with magnetoencephalography (MEG) from healthy young adult subjects while in resting and active states, a systematic framework inspired by thermodynamics is applied to neural EM signals. Despite considerable inter-subject variation in terms of spectral entropy and energy across time epochs, data support the existence of a robust and linear relationship defining an effective state equation, with higher energy and lower entropy in the resting state compared to active, consistently across subjects.
Mechanisms underlying the emergence of relationships between empirically measured effective state functions are further investigated using a model network of coupled oscillators, suggesting an interplay between noise and coupling strength can account for coherent variation of empirically observed quantities. Taken together, the results show macroscopic neural observables follow a robust, non-trivial conservation rule for energy modulation and information generation.}






\begin{figure}[h!]
 	\centering
 	\includegraphics[scale=.5]{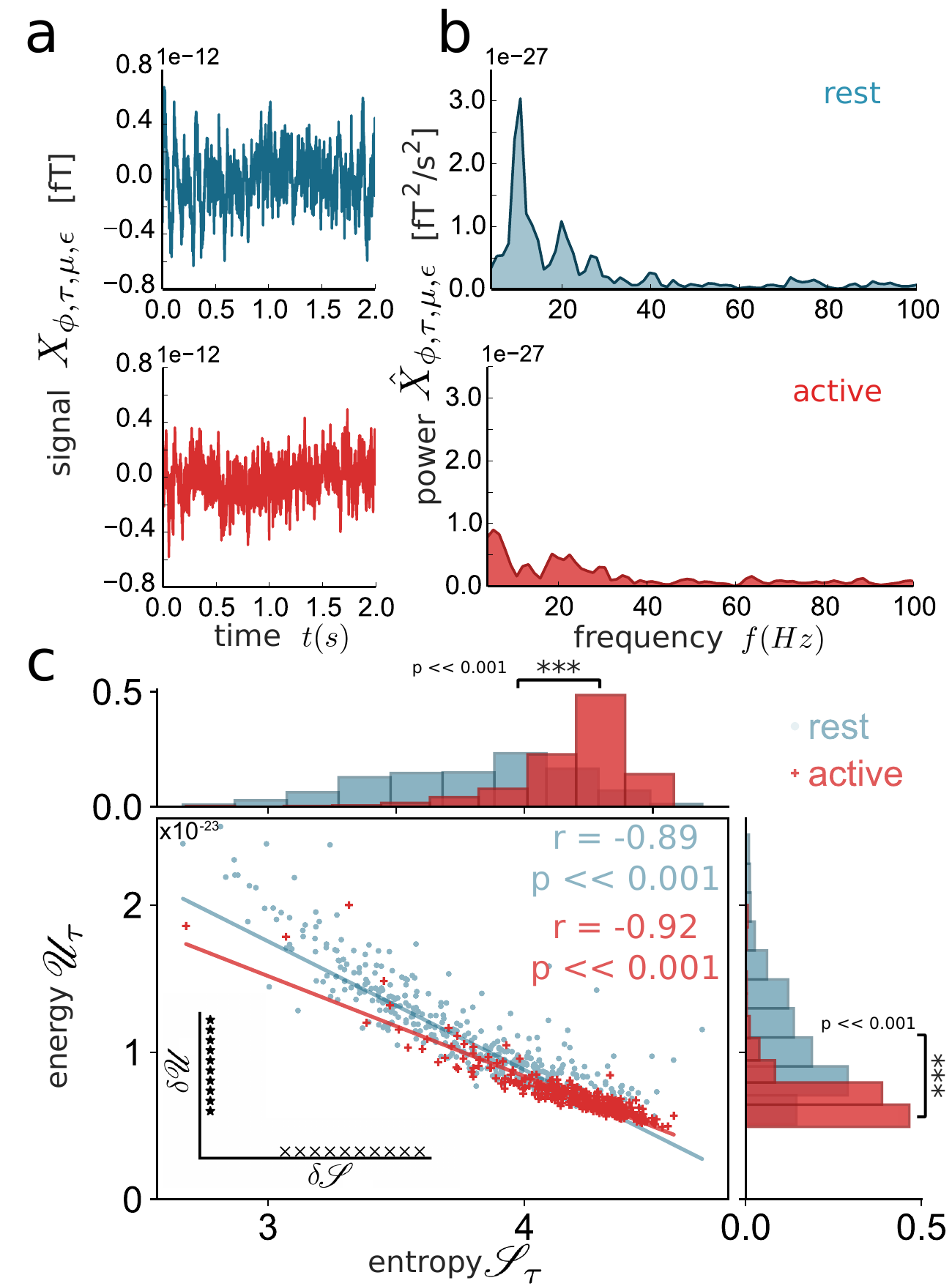}
 	\caption{ \label{fig:pdMF }  \textbf{Lower spectral energy and higher spectral entropy in active versus resting brain state, for one human subject.}  (a) Magentoencephalography (MEG) signals ($X$) from human subjects in resting and active states are used to compute (b) power spectra ($\hat{X}$) for each subject ($\phi$), state ($\tau$), sensor ($\mu$), and epoch ($\epsilon$). (c) Spectral energy (Eq. \ref{energy_def}) and entropy (Eq. \ref{entropy_def}) are derived.  Panel c shows the energy and entropy across epochs for one subject in resting and active states. Between brain states, a significant shift in energy and entropy (Mann-Whitney U test, p-values reported) can be seen in histograms (c) as well as a change in slope and intercept of the negative linear  energy-entropy relation (best fit line shown, Pearson's $r$ reported) as shown in the scatter plot. The inset labeled $\delta \mathscr{S}$, $\delta \mathscr{U}$ implies a non-tautological relation between effective energy and entropy as measured empirically (see text). }
 \end{figure}

Brain activity varies between states, e.g. in unconsciousness vs consciousness, but a formal macroscopic description remains poorly understood. Spectral analysis of neural electromagnetic (EM) signals indicates that fluctuations occur simultaneously at all measurable frequencies, along a 1/f$^\beta$ distribution, above which resonant modes occur at different frequencies \cite{nunez2006electric, niedermeyer2005electroencephalography, buzsaki2006rhythms}. While neural power spectra display broadband features, they are conventionally investigated by \textit{ad hoc} partition of the frequency domain. Results of band-limited analyses have demonstrated that the power of bands shifts between brain states. The shift in power within a frequency band is thought at least in part to reflect changes in the level of synchronous neuronal activity at the time scale of that band. Indeed, vigilant behavioral states are commonly called 'desynchronized' whereas unconscious states 'synchronized' \cite{niedermeyer2005electroencephalography, nunez2006electric, buzsaki2006rhythms}. 

\TA{
Macroscopic, EM brain signals measured outside the human head reflect the coordinated activity of tens of thousands of neurons  \cite{baillet2017magnetoencephalography}. Analogies to statistical physics have helped to understand how microscopic properties of neurons and their interactions lead to neural population behavior at the mesoscale \cite{schneidman2006weak, cocco2009neuronal,marre2009prediction, capone2015inferring, ferrari2016learning, okun2015diverse, gardella2016tractable, capone2018spontaneous} across brain states \cite{tavoni2017functional, nghiem2018maximum}. Since the use of methodology from statistical physics has uncovered descriptions of individual neurons' activity and resulting population behaviour at mesoscopic scales, macroscopically observed changes in signal complexity and magnitude could be addressed in analogy to thermodynamics. In fact, for large-scale, whole-brain recordings, spectral entropy measures reportedly vary with levels of consciousness \cite{sitt2014large, sarasso2015consciousness, mckay2006pharmacokinetic, vanluchene2004spectral}. In the present paper, brain states are described as a function of global state variables, derived from the spectral structure of different human subjects' brain activity.
}

Neural signals from each human subject in each brain state may be dissimilar, but the organization of spectra could follow general principles characterized by a state equation. To test this possibility, a statistical analysis of broadband signals is performed using magnetic fields obtained by magnetoencephalography (MEG) from healthy, adult, human subjects in two brain states - resting state with eyes open and fixated (rest) and while performing an N-back visual working memory task (active) \cite{larson2013adding, van2012human}. 
	
Specifically, the magnetic field X, measured for each subject $\phi=\{1 \ldots 77\}$, brain state $\tau=\{rest, active\}$, sensor $\mu=\{1 \ldots 248\}$, and time epoch $\epsilon=\{1 \ldots M\}$ where $M$ is the total number of epochs (Fig.1) is analyzed. In figure 1a, the time courses of fields for such $X_{\phi,\tau,\mu,\epsilon}\TA{(t)}$ in rest ($\tau=r$, Fig.1a, top) and active ($\tau=a$, Fig.1a, bottom) states is shown. The corresponding power spectra $\hat{X}_{\phi,\tau,\mu,\epsilon}\TA{(f)}$ are computed for each epoch (Fig.1b). 

Observables between resting and active states may present different statistical properties. In order to test this hypothesis, a spectral energy measure $\mathscr{U}_{\phi,\tau,\epsilon}$ is introduced for each subject in each brain state \TA{and each time epoch, summing over sensors :}
\begin{equation}
\mathscr{U}_{\phi, \tau, \epsilon} \equiv \sum_{\mu,f} \hat{X}_{\phi,\tau,\mu,\epsilon}(f)
\label{energy_def}
\end{equation}

\TA{Similarly, a spectral entropy measure $\mathscr{S}_{\phi,\tau,\epsilon}$ is defined, for each subject in each brain state and epoch, summing over sensors, as}

\begin{equation}
\mathscr{S}_{\phi, \tau, \epsilon} \equiv - \sum_{\mu,f} P(f)\,\ln\,P(f),
\label{entropy_def}
\end{equation}
where, \TA{$P(f) := \frac{\hat{X}_{\phi,\tau,\mu,\epsilon}(f)}{\sum_f \hat{X}_{\phi,\tau,\mu,\epsilon}(f)} $} describes the relative heights of power spectral frequency bins, from each subject $\phi$ in each brain state $\tau$.

\begin{figure}
 	\centering
 	\includegraphics[scale=0.5]{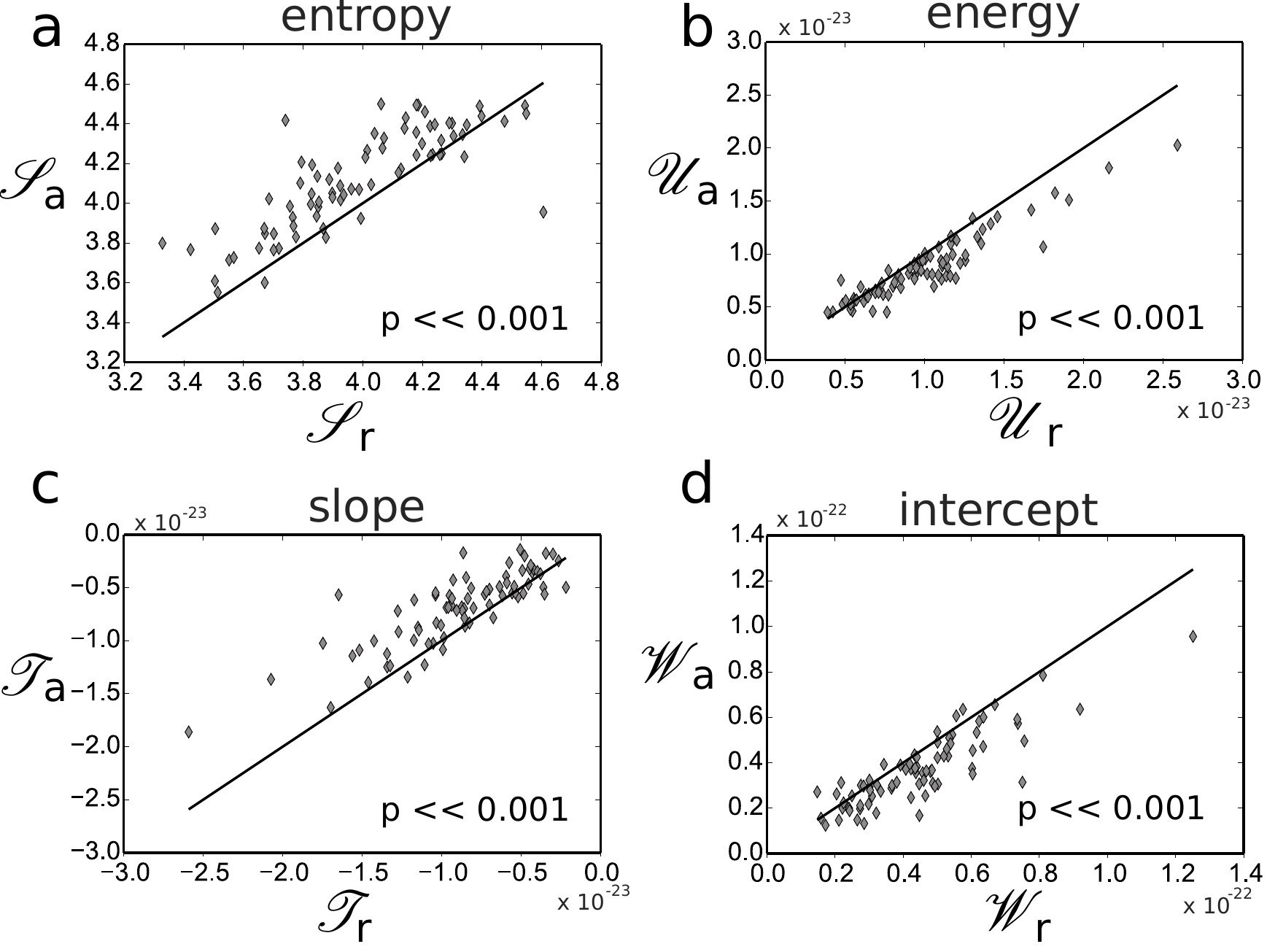}
 	\caption{ \label{fig:pdMF}  \textbf{Variation of the spectral energy-entropy relation between brain states is robust across subjects}. Mean spectral entropy (a), mean spectral energy (b), slope (c), and intercept (d) (Eq. 3) are shown in active state against rest (one point per subject, identity line shown). In the resting state, spectral energy is consistently larger, and entropy smaller than in the active state (Wilcoxon signed-rank test). The energy-entropy relation has a steeper negative slope (c) and a higher intercept (d) in resting than active state.}
 \end{figure}
 
A negative relation between energy and entropy across time epochs is found (Fig.1c): when energy decreases, entropy increases, suggesting a linear relationship of the following form:
\begin{equation}
\mathscr{U}_{\phi, \tau, \epsilon} =  \mathscr{T}_{\phi, \tau}\mathscr{S}_{\phi, \tau, \epsilon} + \mathscr{W}_{\phi, \tau}
\end{equation}
where $\mathscr{T}_{\phi, \tau}$ and $\mathscr{W}_{\phi, \tau}$ are respectively the slope and intercept of the entropy-energy relation across time epochs, for each subject in each brain state. $\mathscr{T}_{\phi, \tau}$ and $\mathscr{W}_{\phi, \tau}$ differ between brain states; in the resting state, energy is higher, entropy is lower, and their relationship follows a steeper (more negative slope) line with a higher-intercept than in the active state (Fig. 1c). 

Such a clear linear relationship between the state variables $\mathscr{S}_{\phi,\tau}$ and $\mathscr{U}_{\phi,\tau}$ raises naturally the question of whether it could be a direct consequence of how the quantities are defined. This possibility is tested through manipulation of the signals where the entropy can be modified ($\delta S$) by adding zero-mean noise to the power spectrum, with no consequent effect on energy as reported in the inset of Fig.1c. Conversely, multiplying the full spectrum by a constant results in a change of the energy ($\delta U$) without altering the entropy. The results support the hypothesis that the observed linear relationship between energy and entropy could reflect fundamental rules that define the way neural activity is organized.

Despite important inter-subject variability in the means over epochs of $\mathscr{U}_{\phi, \tau, \epsilon}$ and $\mathscr{S}_{\phi, \tau, \epsilon}$, as well as $\mathscr{T}_{\phi, \tau}$, and $\mathscr{W}_{\phi, \tau}$, it is verified that the active-rest differences are strongly consistent across subjects, evidence for robustness of the introduced framework (Fig. 2).

Toward a better understanding of mechanisms governing fluctuating oscillations in neural assemblies between brain states, a classical model of coupled oscillators (Kuramoto model) is next studied \cite{kuramoto1975self, breakspear2010generative}. N oscillators of phase $\theta_i, i \in {1,..., N}$ are arranged over a fully connected network, each one evolving according to the following equation: 
\begin{equation}
\dot{\theta_i} = \omega_i + K\sum_{j \neq i}sin(\theta_j - \theta_i) + \sigma \eta_i,
\end{equation}
where $K$ is the coupling strength. The bare frequencies $\omega_i$ are extracted from a bimodal distribution (sum of two Gaussians). A zero-mean white i.i.d. noise term $\eta_i$ of standard deviation $\sigma$ is received by each oscillator.

From simulated network activity, a collective time-varying variable $R$, to compare with $X_{\phi,\tau,\mu,\epsilon}$, is defined:
\begin{equation}
R = r\frac{1 - \cos(\Theta)}{2},
\end{equation}
where 
\begin{equation}
re^{i\Theta} = \sum_{j}e^{i\theta_j}.
\end{equation}
The parameter $r(t)$ captures the synchronization between the oscillators - indeed, when the phases are equally distributed over $[0, 2\pi]$, $r(t) = 0$. Conversly, when the system is completely synchronous, all oscillators have the same phase, so that $r(t) = 1$. The variable $\Theta(t)$ represents the time evolution of the system and describes the frequency of synchronous events.

To investigate further the behavior of the effective energy-entropy relation found in human data, the dynamics of the Kuramoto model is investigated by changing the coupling parameter $K$ and the amplitude of noise $\sigma$, which are known to strongly modulate synchrony and complexity in coupled oscillator models \cite{pikovsky2003synchronization}. Varying $K$, a transition from asynchronous active-like to synchronous rest-like dynamical regime (Fig.3a-c) occurs. To mimic the observed variation between time epochs, the amplitude of noise $\sigma$ can be understood as varying in time (Fig.3d-f). Increasing noise amplitude $\sigma$ results in less synchronous activity, consistent with previous reports, as well as decreased $R$ \cite{pikovsky2003synchronization}, reduced energy, and increased entropy (Fig. 3). Altogether, the results indicate that a change in brain state can be modeled by varying coupling strength, while variation across time can be simulated by fluctuations in noise amplitude. 

\begin{figure}
 	\centering
 	\includegraphics[scale=0.4]{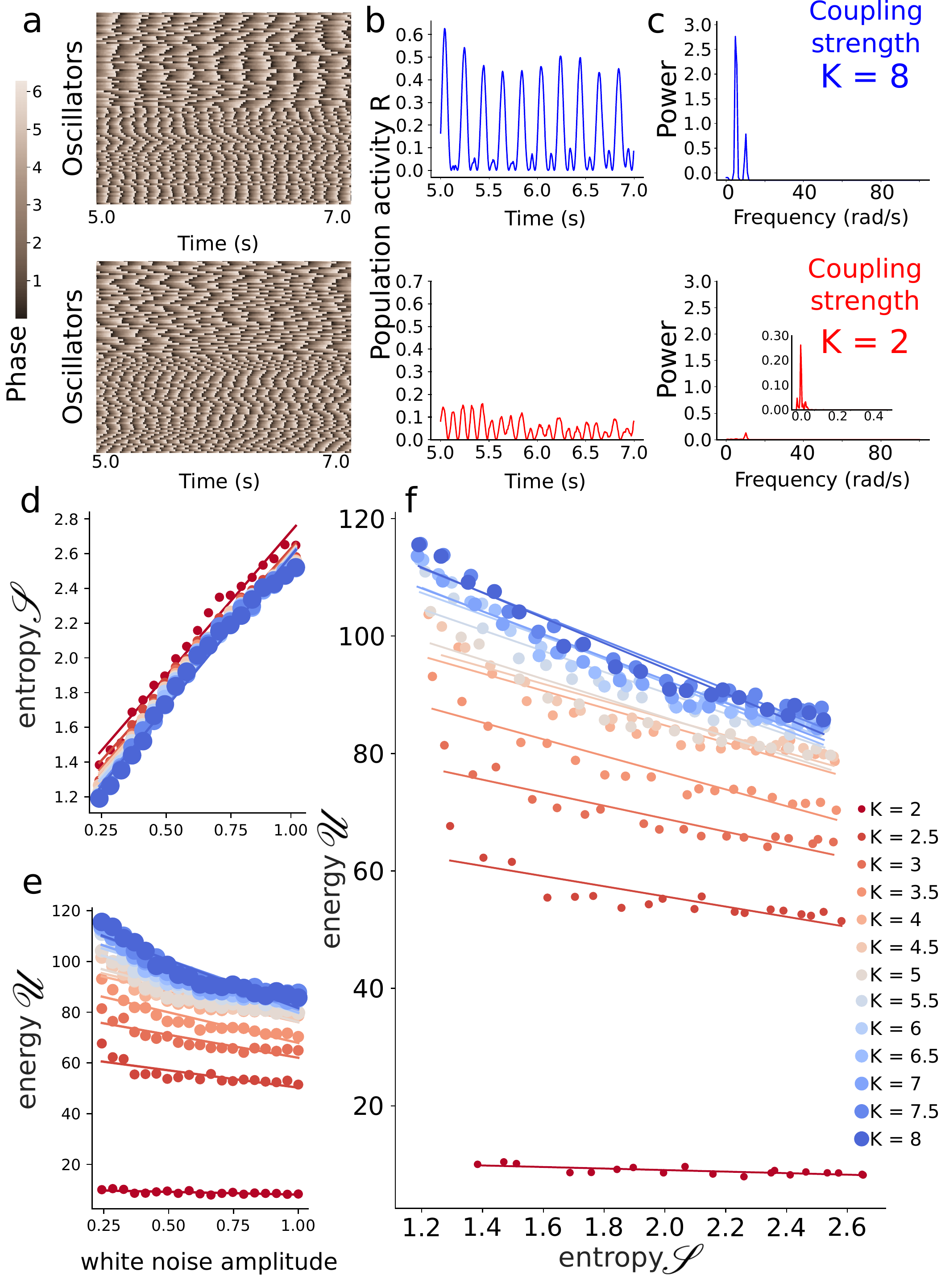}
 	\caption{ \label{fig:Kuramoto}  \textbf{Phase coupling and noise modulate spectral energy and entropy.} A parameter exploration of the Kuramoto model is performed in order to explain variation of state variables in experiments. Change in coupling strength $K$ between oscillators causes a shift in spectral state variables. (a) Phase of each oscillator throughout time. (b)  R (Eq. 5). (c) Power spectra of R. (d-e) To reproduce variation across epochs, noise amplitude is modulated - the higher the noise, the larger the spectral entropy and the lower the spectral energy. (f) Energy and entropy as defined therefore vary along a negative linear relation, whose slope becomes steeper and intercept increases with coupling strength.}
 \end{figure}

\begin{figure}
 	\centering
 	\includegraphics[scale=0.4]{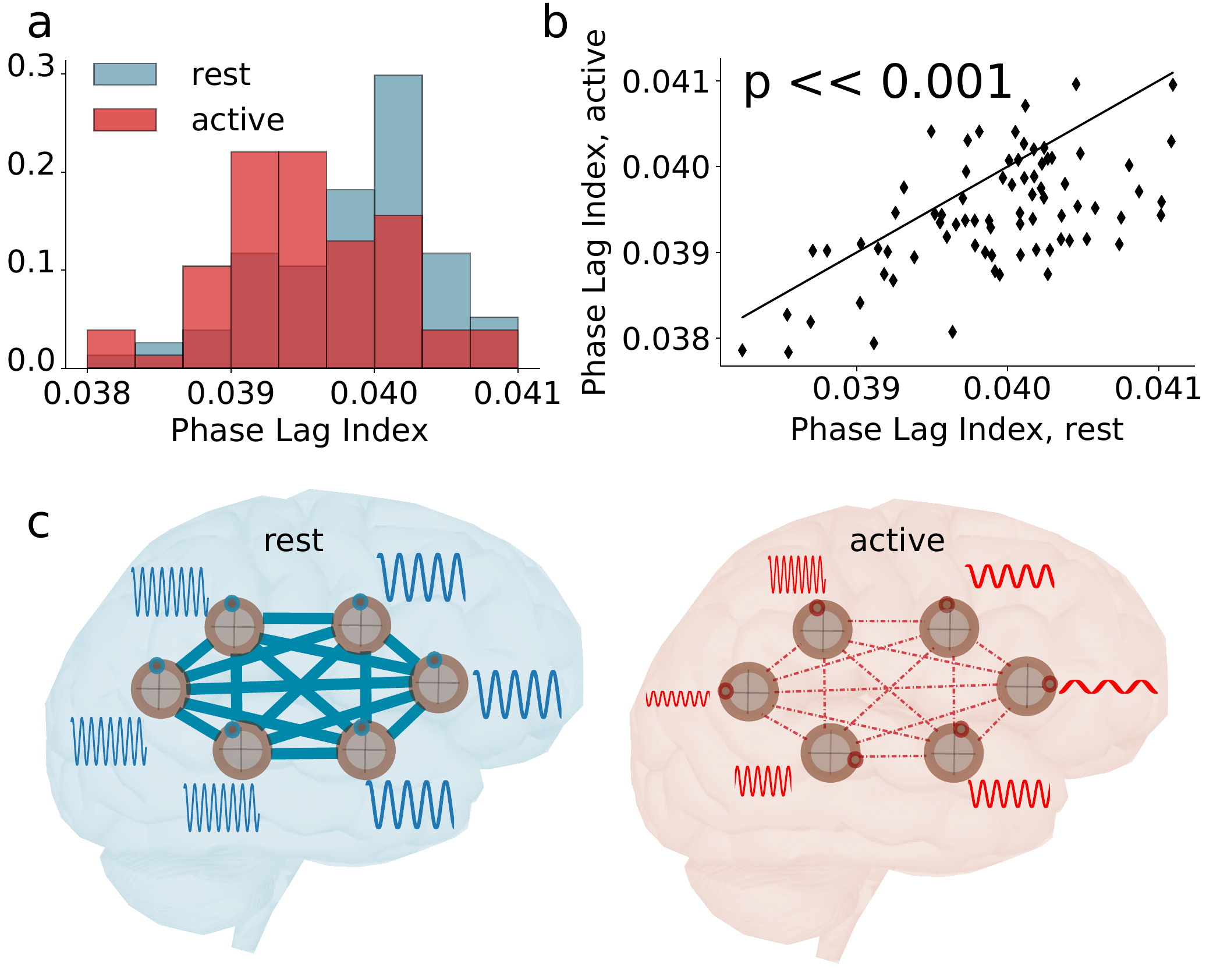}
 	\caption{ \label{fig:model_prediction}  \textbf{Phase Lag Index (PLI) is larger in resting than active state, consistent with stronger coupling during rest.} The  PLI is computed across frequency bins in each epoch in rest and task for each subject. (a-b) The mean PLI across epochs for each subject and state is reported, showing the PLI is significantly larger during the resting than active state (paired Student's T-test). (c) Summary diagram of the results. The data support a model in which variation of effective energy and entropy between brain states relies on noise fluctuations and a shift in coupling strength.}
 \end{figure}

If changes in coupling contribute to empirically observed shifts in neural state variables as predicted by the model, one may expect to find alterations in the phase synchrony in the data. In order to test this prediction, the Phase-Lag Index (PLI), a measure of synchronization across time series is applied to human data \cite{silva2017effect}. By definition, the PLI is larger in more phase-locked states. Indeed, in resting compared to active state data, a significant trend toward enhanced phase synchrony is found (Fig.\ref{fig:model_prediction}). 

A potential interpretation of the robust relation between these effective state variables could reflect the fact that energy is used to generate information in brain activity. It is important to note that the energy quantity proposed here is an indirect measure of neural activity and may be non-trivially related to actual internal energy. We also find evidence for a change in coupling, due to the consistent change in intercept and slope of the energy-entropy relation. Coupling regulates the sensitivity of networks to noise, which can account for the slope of the aforementioned relation. Indeed, with strong coupling, perturbations propagate more efficiently throughout the network, causing changes in the collective activity and affecting the resulting energy. Conversely, in a weakly-coupled network, perturbations do not easily spread across the system. In other words, the 'gas-like' regime found in active brains may be useful for the potential to generate information versus the 'liquid-like' state in which strong coupling promotes a low entropy state with less informational capacity. This provides a new interpretation of previously reported changes in functional connectivity across brain states \cite{biswal1995functional, buckner2008brain, raichle2001default, massimini2005breakdown, bettinardi2015gradual}. 

Lastly, in light of statistical physics inspired models of neuronal activity \cite{schneidman2006weak, cocco2009neuronal,marre2009prediction, capone2015inferring, ferrari2016learning, okun2015diverse, gardella2016tractable, tavoni2017functional, nghiem2018maximum, capone2018spontaneous}, our results suggest a strategy to bridge the gap from macroscopic system signals studied here to microscopic cellular level activity. Indeed, biophysical models of spiking neural networks could be used to investigate how the mechanisms uncovered with the Kuramoto model translate at the microscopic scale. For instance, biophysical processes that control the variability of noisy inputs or the network's tendency to synchronize could be studied. This may also provide insight on how the spectral state variables defined here relate to entropy and energy of micro-states in the sense of statistical physics and spin glass models. While such an approach demands scale-integrated data from a sufficiently large population of single neurons simultaneous with global electromagnetic measurements, it would pave the way to more formal, scale-integrated descriptions of brain activity in healthy and pathological states.
\section{Acknowledgements}
Data were provided by the Human Connectome Project, WU-Minn Consortium (Principal Investigators: David Van Essen and Kamil Ugurbil; 1U54MH091657) funded by the 16 NIH Institutes and Centers that support the NIH Blueprint for Neuroscience Research; and by the McDonnell Center for Systems Neuroscience at Washington University. This project was supported by the European Union (Human Brain
Project H2020-720270 and H2020-785907). JSG was supported by the Ludmer Centre for Neuroinformatics and Mental Health, the Molson Neuroengineering Award, The NSERC-CREATE Training Program in Neuroengineering, and the Ann and Richard Sievers Foundation for Innovation in Neuroscience. JML was supported by an NSERC Discovery grant. The authors thank Gleb Bezgin, Bruno Cessac, Alexis Garcia, Peter Grutter, George Kostopoulos, Naj Mahani, Danko Nikolic, Bartosz Tele\'nczuk, and N\'uria Tort-Colet  for useful discussion. \section{Author contributions}
T.-A. N. wrote data analysis codes, investigated the Kuramoto model, contributed to interpreting the results, and writing the manuscript.
J.-M. L. contributed to writing the analysis codes and the manuscript.  
M. d. V. contributed to interpreting the results, investigating the Kuramoto model, and writing the manuscript. 
C. C. investigated the Kuramoto model, contributed to interpreting the results and editing the manuscript.
A. C. E. provided resources and supervision of the work.
A. D. designed research, provided supervision, and edited the manuscript. 
J. S. G. conceived of the question, designed research, analyzed data, interpreted results, and wrote the paper. 
\section{Competing interests}
The authors declare no competing interests.
\section*{References}
%

\newpage
\section*{Supplementary information}
\subsection*{Data and preprocessing}
Magnetoencephalography (MEG) data were acquired, preprocessed, anonymized, and distributed by the Human Connectome Project (HCP). Healthy adult human subjects were imaged in a magnetically shielded room using a whole head MAGNES 3600 (4D Neuroimaging, San Diego, CA) system at the Saint Louis University (SLU) medical campus. HCP protocols collected resting state (rest) and visual N-back working memory (active) data from the same subjects. The MEG has 248 magnetometer channels. Electrooculography, electrocardiography, and electromyography recordings are all synchronized with the MEG. Data were acquired with a bandwidth of 400Hz, DC high pass filter, in a continuous acquisition mode, with a 2034.5 Hz sampling rate. MEG data were preprocessed by the HCP. Biological and environmental artifacts, including 60 Hz artifacts, bad channels, and bad segments were removed first automatically by an independent component analysis (ICA)-based, publically available, custom Fieldtrip-scripted pipeline and further quality controlled by human verification. Channel-level preprocessing pipelines also downsampled data to 508.63Hz. N-back data were split into groups of trials corresponding to task design, and divided into epochs time-locked to the onset of the stimulus (TIM) to yield Matlab structures containing epoched, preprocessed data. Further documentation describing data collection and preprocessing is available \cite{larson2013adding} and
$www.humanconnectome.org/documentation$.
 
\subsection*{Power spectrum} 
Subjects with data in both N-back memory and resting states (N = 77) were included in the analyses. Per subject average recording time is:  14.06$\pm$0.55 min for resting state and  12.35$\pm$0.67 min for working memory. Respecting constraints introduced by epoch time and Nyquist artifacts, power spectra between [4,100] Hz were obtained with Welch windowing method per $X_{\phi,\tau,\mu,\epsilon}$, indexing 2 sec MEG time series from each subject $\phi$, brain state $\tau$  (active or rest), sensor $\mu$, and epoch $\epsilon$. 

\subsection*{Spectral energy and entropy}
Each power spectrum $\hat{X}_{\phi, \tau, \mu, \epsilon}(f)$ was used to construct a histogram with bin size 0.1 Hz. Summing over the histogram bins of the power spectra of all sensors, the energy $\mathscr{U}_{\phi, \tau, \epsilon}$ is obtained (Eq. \ref{energy_def}). By normalizing each histogram, one obtains the frequency probability distribution $P(f)$ for each sensor, epoch, state, and subject. Once again, summing over all bins and all sensors, the entropy can be derived following Eq. \ref{entropy_def}.

\subsection*{Kuramoto model}
In all our simulations, we consider an all-to-all network of $N = 100$ oscillators. The bare frequencies of each oscillator $\omega_i$ are extracted from a distribution of the following form:
\begin{equation}
	p(\omega) = \frac{1}{\sqrt{8\pi\Delta}}\Big\{\exp\Big[-\frac{(\omega - \omega_0)^2}{2\Delta^2}\Big] + \exp\Big[-\frac{(\omega - \omega_1)^2}{2\Delta^2}\Big]\Big\},
\end{equation}
where $\Delta = 1$ Hz, $\omega_0 = 5$ Hz, and $\omega_1 = 10$ Hz. We verified that the general picture reported in the main text is not affected by the specific choice of these three parameters. 
The simulations were run for 15 minutes of simulated time (to match with the experimental data and to ensure a robust estimation of observables), using an Euler integration method with a time step of 5 ms. The energy and entropy are computed on power spectra estimated on windows of 2 s, as with the experiments. A transient time of 100 s was discarded in all simulations. 

\subsection*{Phase Lag Index (PLI)}
The PLI was computed for each subject and time window, summing over all sensors. The frequency range of interest [4,100] Hz was divided into ten equally wide bins. Within each frequency bin, a time series is obtained by filtering the signal with a Butterworth band-pass filter. The Hilbert transform is then employed to extract the phase $\psi_k$ of the time series in each frequency bin $k$. From there, the PLI, given by
\begin{equation}
    \mathrm{PLI} \equiv \lvert <sign(\psi_k)>\rvert,
\end{equation}
is computed, where $<\boldsymbol{\cdot}>$ denotes averaging over time \cite{silva2017effect}. One may note that the PLI takes values between 0 (random phase relations) and 1 (perfect phase locking).
In this work we report the mean PLI over all time epochs for each subject in each brain state.
\end{document}